\documentclass[aps,twocolumn]{revtex4}
\usepackage{epsfig}
\usepackage{amsmath}
\usepackage{epsfig}

\begin{document}
\title
{Coherent scattering from semi-infinite non-Hermitian potentials}    
\author{Zafar Ahmed$^1$, Dona Ghosh$^2$, Sachin Kumar$^3$}
\affiliation{$~^1$Nuclear Physics Division, $~^3$Theoretical Physics Section \\ Bhabha Atomic Research Centre, Mumbai 400 085, India
\\
$~^2$Department of Mathematics, Jadavpur University, Kolkata 700032, India}
\email{1:zahmed@barc.gov.in, 2: rimidonaghosh@gmail.com, 3: sachinv@barc.gov.in}
\date{\today}
\begin{abstract}
When two identical (coherent) beams are injected at a semi-infinite non-Hermitian medium from left and right, we show that
both reflection $(r_L,r_R)$ and transmission $(t_L,t_R)$ amplitudes are non-reciprocal. In a parametric domain, there exists
Spectral Singularity (SS) at a real energy $E=E_*=k_*^2$ and the determinant of the time-reversed two port scattering matrix  i.e., $|\det(S(-k))|=|t_L(-k) t_R(-k)-r_L(-k) r_R(-k)|$
vanishes sharply at $k=k_*$ displaying the phenomenon of Coherent Perfect Absorption (CPA). In the complimentary parametric domain,
the potential becomes either left or right reflectionless at $E=E_z$. But we rule out the existence
of Invisibility despite $r_R(E_i)=0$ and $t_R(E_i)=1$ but $T(E_i) \ne 1$, in this new avenue. We present two simple exactly solvable models where  expressions for $E_*$, $E_z$, $E_i$ and  parametric conditions 
on the potential have been obtained in explicit and simple forms. Earlier, the novel phenomena of SS and CPA have been found to occur only in the scattering complex potentials which are spatially localized (vanish asymptotically) and have $t_L=t_R$.
\end{abstract}	
\maketitle
A non-Hermitian complex potential $V(x)=V_r(x) + iV_i(x)$ which is spatially localized and non-symmetric displays the non-reciprocity of reflection amplitudes ($r_L \ne r_R $) whereas transmission amplitudes are reciprocal $t_L=t_R$ [1-7]. For non-Hermitian scattering potentials
the existence of a special real energy ($E_*$) has been proposed [8]  where all three probabilities ($T=|t|^2, R=|r|^2)$ $T(E), R_L(E)$ and $R_R(E)$ become infinity. This special energy is called Spectral Singularity (SS) [8]. Though SS was first demonstrated to exist in a complex PT(Parity and Time)-symmetric potential [8],  with ample number of examples, later it has been found  [9]   that SS is a property of either a complex  non-PT-symmetric potential or the parametric domain of broken PT-symmetry of a complex PT-symmetric potential. Very interesting exactly solvable models are available [10] where one gets explicit expression of $E_*$ and explicit parametric conditions on the non-Hermitian potential. 

The two concepts: non-reciprocity [1-7] of reflection
 and the spectral singularity [8] give rise to a new experimentation where coherent (identical) beams are injected into a non-Hermitian optical medium  from left and right. In the coherent scattering, the determinant of two port scattering matrix $S(k)$ is given as $|\det(S(k))|=|r_L(k) r_R(k)-t^2(k)|$ [11]. It has  been further claimed that if spectral singularity occurs at $E=E_*=k_*^2$, $S(-k)$ becomes zero at $k=k_*$ signifying perfect absorption of coherent beams  [11] in the non-Hermitian medium. This novel idea of Coherent Perfect Absorption (CPA) has given rise to time reversed Lasers [11-13]. The complex PT-symmetric potentials have been ruled out [9] for CPA which is also referred to as coherent perfect absorption without lasing [11-13].

We would like to remark that unlike the first proposal for the general CPA [11], the authors in [14] have been cautious about choosing the optical non-Hermitian medium as Parity-symmetric. 
They set less general, yet simpler and intuitive condition for CPA at a real energy as $t+r_{L}=0=t+r_{R}$. For Parity-symmetric complex potentials the reciprocity ($r_{L}=r_{R}$) works. This phenomenon has been called controlled CPA which is a special case of the more general condition [11]. The existence of SS in this case also supports the conjecture that complex PT-symmetry  is not necessary for SS. Very interesting exactly solvable models of CPA have been proposed [10]. For the non-Hermitian PT-symmetric potentials which are spatially localized, the Unidirectional Invisibility (UI) [15,16] occurs when either $r_L$ or $r_R$ vanishes at a real energy $E=E_i$ and  $t_L=t_R=t=1$ at this energy. For complex PT-symmetric potentials another novel phenomenon of CPA with lasing has been revealed [17,18]. Very interestingly aforementioned phenomena occur as a possibility and not as a necessity, so their (non) occurrence  in various systems are worth studying. For instance, recently, (non) occurrence of SS has been discussed [19] in terms of various kinds of anti-linear symmetry of the spatially localized optical mediums. 

In a sharp contrast to the aforementioned works [1-19] on scattering where  complex potentials are spatially localized, in this paper we study scattering from complex potentials where the real part is semi-infinite and the imaginary part is as usual spatially localized (see Fig. 1).
\begin{figure}[h]
\centering
\includegraphics[width=7 cm,height=5 cm]{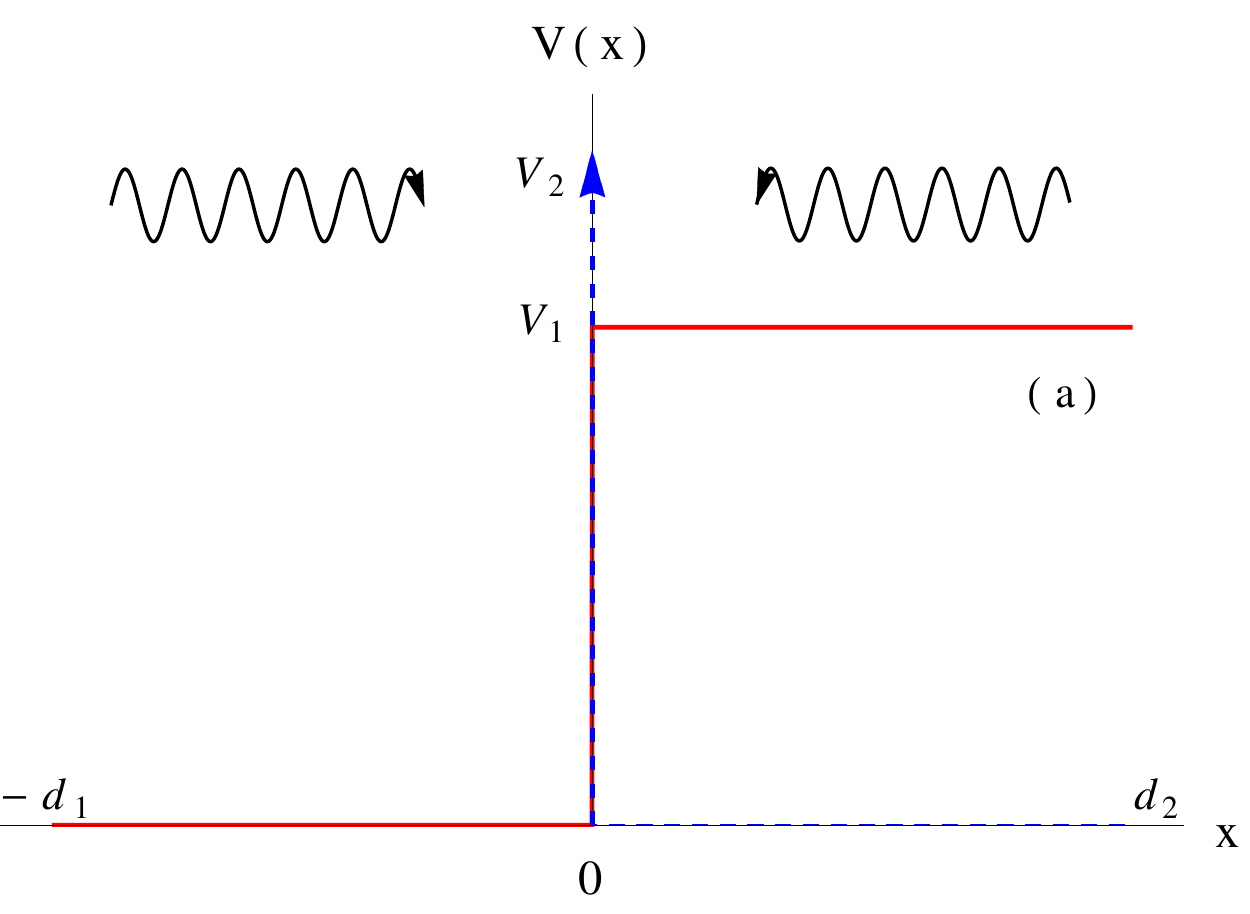}
\hspace{.5cm}
\includegraphics[width=7 cm,height=5 cm]{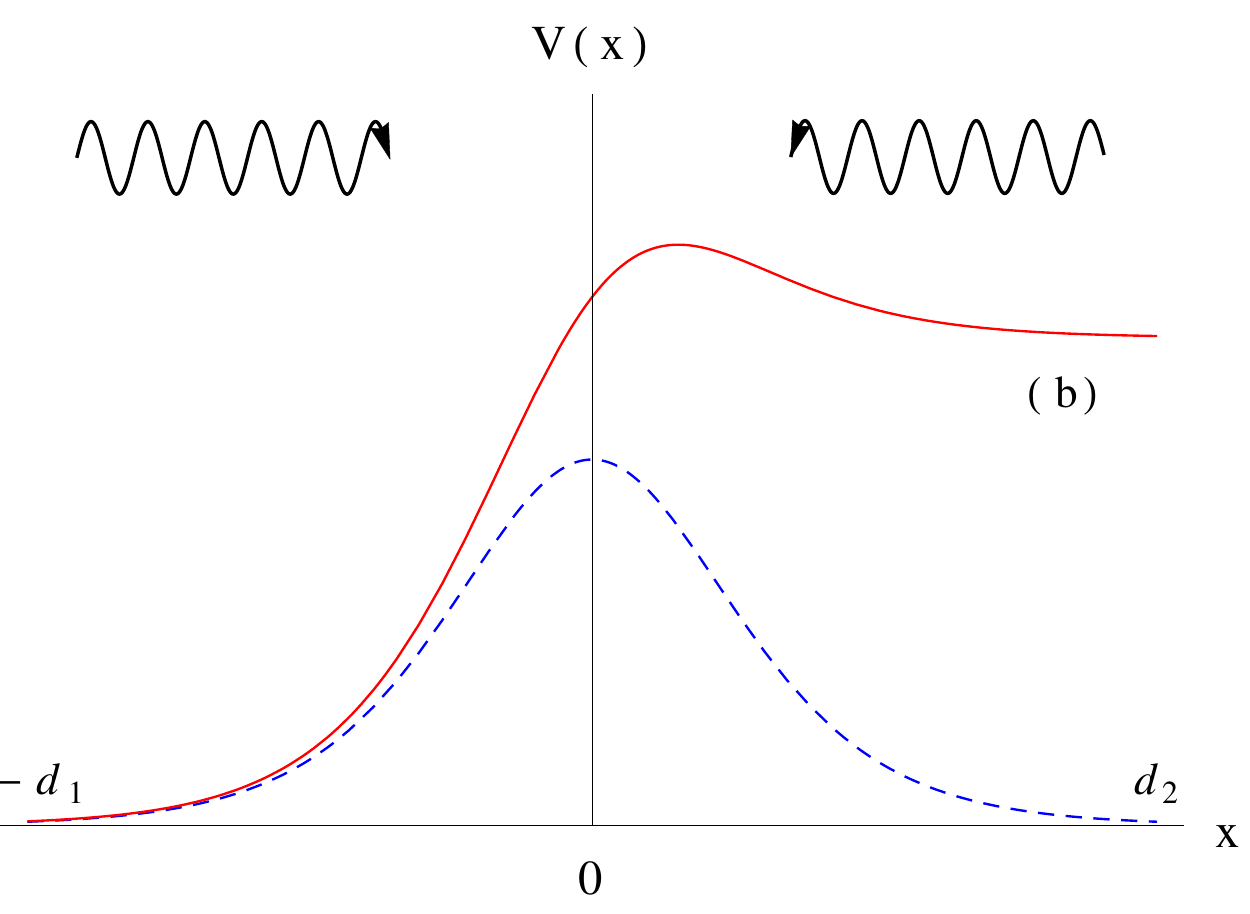}
\caption{Two models of non-Hermitian  potentials for coherent scattering from left and right.  Their real part are semi-infinite. (a): Eq. (10), (b): Eq. (11). Solid lines are for real and dashed lines  denote imaginary part. The vertical arrow in (a) represents the Dirac delta potential.}
\end{figure}
By semi-infinite, we mean that $V_r(x\sim -\infty)=0$ and $V_r(x \sim \infty)=V_1$. So the potentials discussed here are essentially non-PT-symmetric. In these interesting models, we find that both $r$ and $t$ are non-reciprocal (see below Eq. (7,8)). Nevertheless, the transmission probabilities remain reciprocal again: $T_L=T_R$ in a non-trivial way (see Eq. (14) below). The question arising here is as to whether we can observe the novel phenomenon of SS, CPA and UI even in such semi-infinite mediums. In this paper, we derive the two-port s-matrix (see Eqs. (7,8) below) for coherent injection at a semi-infinite potential and investigate the possibility of occurrence of the aforementioned phenomena of SS, CPA and UI yet again. Earlier, a non-hermitian semi-infinite potential has been studied however it being one   unit of a periodic array  does not remain semi-infinite, nevertheless, this study gives rise to several other interesting issues [20] in scattering from optical potentials.

The Schr{\"o}dinger equation for a semi-infinite potential $V(x)$ (see Fig. (1))
\begin{equation}
\frac{d^2\psi(x)}{dx^2}+\frac{2\mu}{\hbar^2}[E-V(x)] \psi(x)=0
\end{equation}
is solved by defining $k_L{=}\sqrt{2\mu E}/\hbar, k_R{=}\sqrt{2\mu(E-V_1)}/\hbar.$ 
Let $d_1$ be large asymptotic distance such that $V(-d_1)=0$ and $V(d_2)=V_1$. $u(x)$ and $v(x)$ are two linearly independent solutions of Schr{\"o}dinger equation in the interval $[-d_1,d_2]$ such that $u(0)=1, u'(0)=0; v(0)=0, v'(0)=1$. These conditions ensure linear independence of the
two solutions of the second order differential equation (1) and the constancy (independence on $x$) of the Wronskian: $W(x)=[u(x)v'(x)-u'(x)v(x)]$ for all $x$ $ \in (-\infty, \infty)$. The solution of (1) for the semi-infinite models is given by
\begin{eqnarray}
\psi(x<-d_1)\nonumber = A_L e^{ik_Lx} +B_L e^{-ik_Lx} \\  \psi(-d_1<x<d_2) \nonumber = C u(x) + D v(x) \\  \psi(x>d_2)  =A_R e^{ik_Rx} + B_R e^{-ik_Rx}.
\end{eqnarray}
Next the numerical integration on both sides provides us with the end values $u(-d_1), u'(-d_1), v(-d_1), v'(-d_1)$ at a given energy $E$ on the left of the potential. For short we will denote these values as $u_1, u'_1, v_1, v'_1$; respectively. Similarly, we will have
$u_2, u'_2, v_2, v'_2$ evaluated at $x=d_2$. The quantities $u_1,v_1,u_2,v_2$ are in general complex. 

Further, we use the transfer matrix method [8,18,21] of scattering in one-dimension.
Matching these solutions and their first derivative at $x=-d_1$ and $x=d_2$, in the matrix notation we can write
\begin{multline}
\begin{pmatrix} 
A_L \\ B_L     
\end{pmatrix} 
= 
\begin{pmatrix}
 f^{-1} & f \\   
ik_L f^{-1} &-ik_L f  
\end{pmatrix}^{-1}  
\begin{pmatrix}   
u_1 & v_1 \\       
u'_1 & v'_1 
\end{pmatrix} 
\begin{pmatrix}  
 u_2 & v_2 \\ 
u'_2  & v'_2   
\end{pmatrix}^{-1} \\                                  
\begin{pmatrix}   
 g & g^{-1} \\ 
ik_{R} g &-ik_{R} g^{-1} \\                                                          
\end{pmatrix}   
\begin{pmatrix}  
 A_R \\ B_R   
\end{pmatrix}  
\end{multline}
where $f=e^{ik_L d_1}$ and $g=e^{ik_R d_2}$.
For short the matrix product can be denoted as $M=M_1^{-1}M_2 M_3^{-1} M_4$, which is called transfer matrix. The Wronskians $u_1v'_1-u'_1v_1=W{=}u_2 v'_2-u'_2v_2$ give us $\det(M){=}k_R/k_L$.
When $k_L{=}k_R$, we get $\det(M){=}1$. Let us point out that normally $\det(M){=}1$ is used as fundamental property of the transfer matrix wherein the crucial and more basic connection of Wronskian is often  overlooked. We denote the  product $M_2M_3^{-1}$ as $M_5$ to write
\begin{multline}
M_5      
= 
\begin{pmatrix}
u_1 v'_2-v_1 u'_2 & u_2 v_1-u_1 v_2 \\ 
u'_1 v'_2-u'_2 v'_1 & u_2 v'_1-u'_1 v_2 
\end{pmatrix} =
\begin{pmatrix}  
 w_{11} & w_{12} \\ 
 w_{21} & w_{22}
\end{pmatrix} 
\end{multline}
$\det(M_5)=1$ holds once again.
Now the transfer matrix $M$ can be denoted as
\begin{equation}
M=M_1^{-1} M_5 M_4= \left ( \begin{array}{cc} m_{11} & m_{12} \\ m_{21} & m_{22} \end{array} \right).
\end{equation}
From (3,5), we get $A_L =m_{11} A_R+ m_{12} B_R$ and $B_L=m_{21} A_R +m_{22} B_R$. So we can write

\begin{figure}[h]
	\centering
	\includegraphics[width=7 cm,height=5 cm]{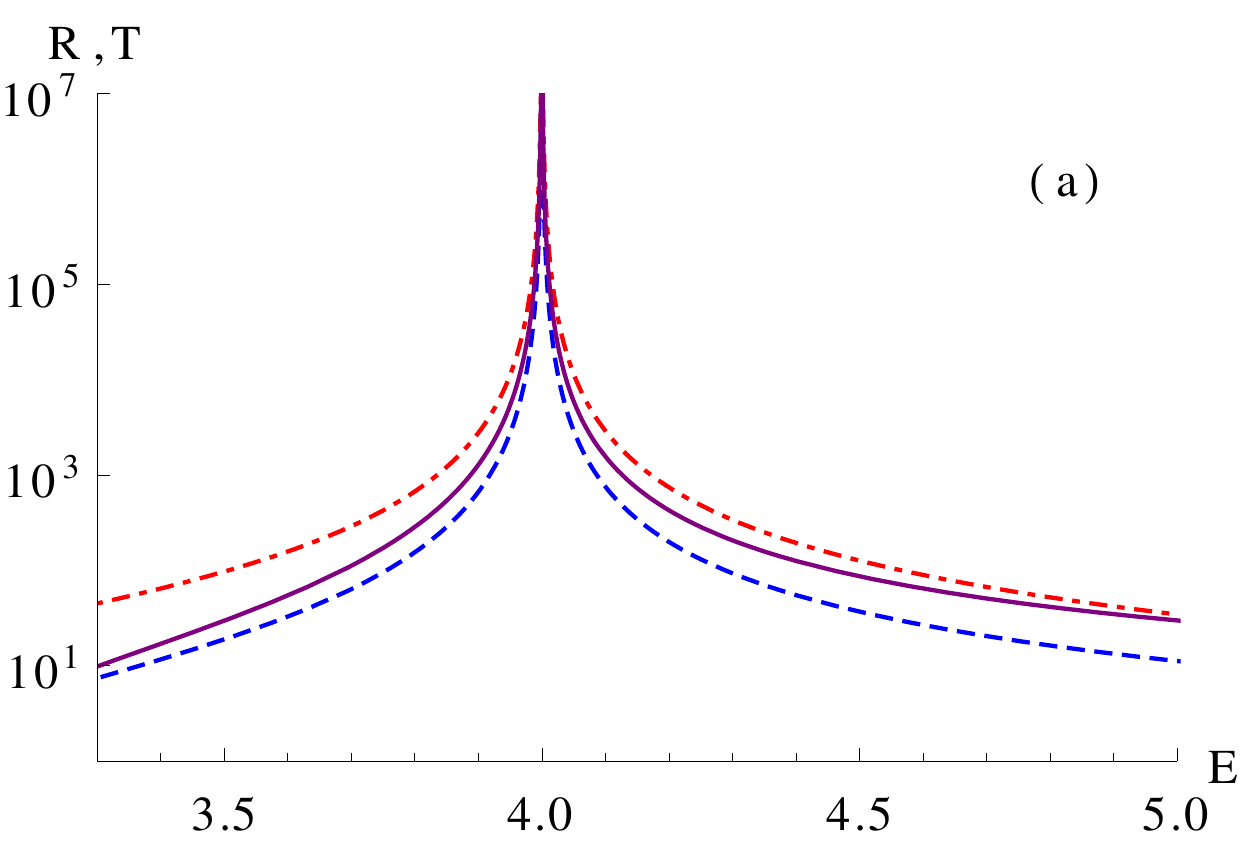}
	\hspace{.5cm}
	\includegraphics[width=7 cm,height=5 cm]{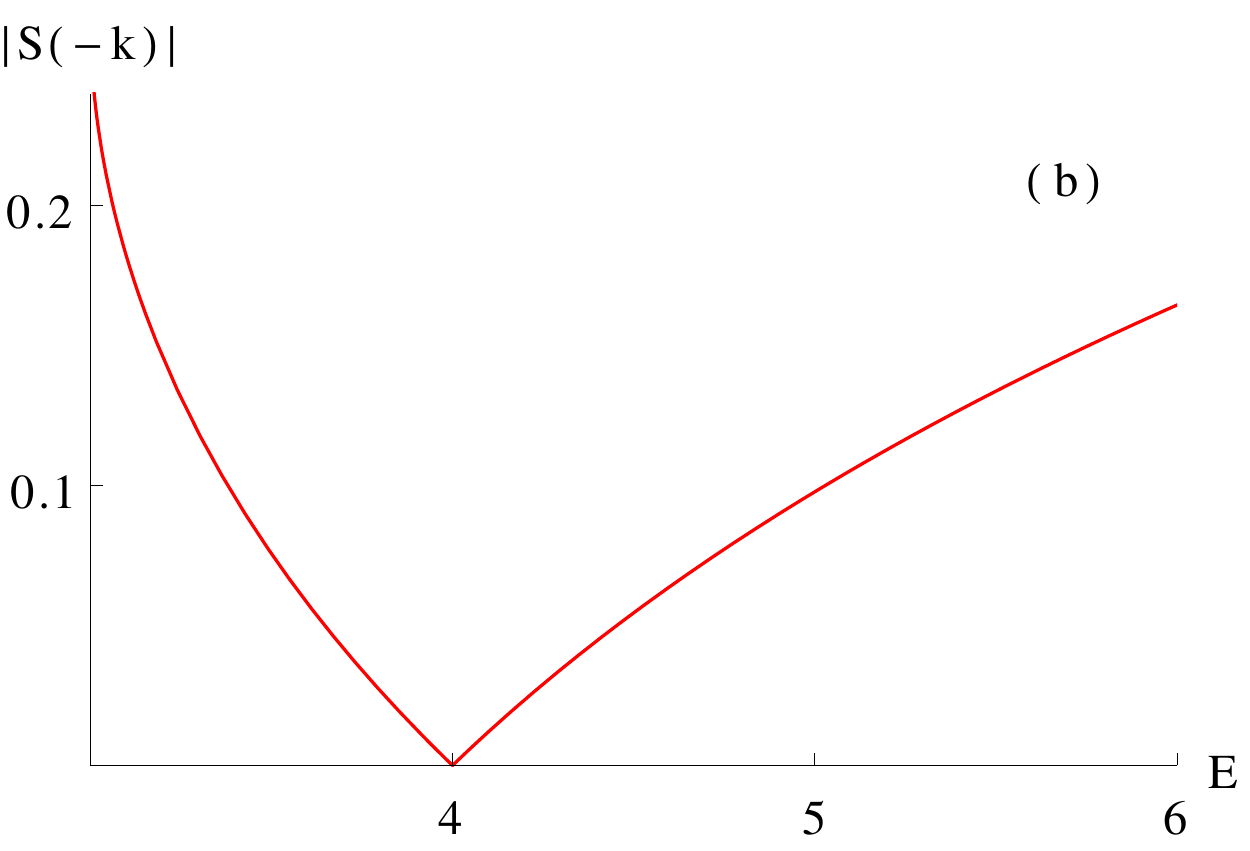}
	\caption{(a):$T(E), R_L(E)$ and $R_R(E)$ displaying spectral singularity at $E=4$, for the model (10) of semi-infinite non-Hermitian potential.
	In (b): see $|\det(S(-k_L,-k_R))|$ passing sharply through zero at $E=4.$ Here, $2\mu=1=\hbar^2, V_1=3, V_2=3$.}
\end{figure}
\begin{equation}
\left(\begin{array}{c} A_R \\ B_L \end{array} \right) = \left ( \begin{array}{cc} -m_{11} & 0 \\ -m_{21} & 1 \end{array} \right) ^{-1} \left ( \begin{array}{cc} -1 & m_{12} \\ 0 & m_{22} \end{array} \right) \left(\begin{array}{c} A_L \\ B_R \end{array} \right).
\end{equation}
Hence the two port S matrix $S(E)$ for coherent injection from left and right gets defined as
\begin{multline}
S(k)=
\begin{pmatrix} 
s_{11} & s_{12} \\ 
s_{21} & s_{22}     
\end{pmatrix} 
= 
\begin{pmatrix}
\frac{1}{m_{11}} & -\frac{m_{12}}{m_{11}} \\
 \frac{m_{21}}{m_{11}} & \frac{\det(M)}{m_{11}}
\end{pmatrix}=
\begin{pmatrix}   
t_L & r_L \\ 
r_R & t_R
\end{pmatrix} 
\end{multline}
We also prove the non-reciprocity of transmission amplitude for semi-infinite potentials  as
\begin{equation}
k_L t_R=k_R t_L.
\end{equation}
The crucial question arising here is as to whether a semi-infinite potential $V_r(x)$ (${\cal R}(V(x))$) can also give rise to CPA with a modified two-port S-matrix (7), where
\begin{equation} 
 |\det(S(k_L,k_R))|=|t_L t_R -r_L r_R|.
\end{equation}
For this we propose two semi-infinite models. The first one has sharp semi-infinite potential step as real part and Dirac delta function as imaginary part:
\begin{equation}
V(x){=}V_1 \Theta(x) {+} iV_2 \delta(x),~\Theta(x{<}0){=}0,~ \Theta(x{>}0){=}1.
\end{equation}
The other one has its real part as defused Fermi step  and imaginary part as $\mbox{sech}^2 x$:
\begin{equation}
V(x)=\frac{V_1}{2}[1+\tanh (\frac{x}{2a})]+ i V_2 \mbox{sech}^2(\frac{x}{2a}).
\end{equation}
In (10,11), $V_1$ is essentially real and $V_2$ may be non-real such that ${\cal R}(V_2)>0$. 
 The Hermitian version of the potential (11) is well known as Eckart [22] or Rosen Morse [23] potential which are known to be exactly solvable. More recently a complex PT-symmetric version of the Eckart or Rosen-Morse potential: $V(x){=}A \mbox{sech}^2 x+i B\tanh x$ ($A,B$ are real) has been studied for scattering [24]. However, here, we utilize it as an essentially non-PT-symmetric complex non-Hermitian potential.

For the scattering from left  for (10) we take $\psi(x{<}0){=} A_L \exp(ik_Lx){+}B_L \exp(-ik_Lx), ~~~ \psi(x{>}0){=}A_R \\ \exp(ik_Rx)+ B_R \exp(-ik_Rx)$. By matching the solutions at $x=0$ and  mismatching their first derivative at $x=0$ due to the presence of Dirac delta potential, we obtain  
\begin{equation}
r_L{=}\frac{B_L}{A_L}{=}\frac{k_L-k_R+u}{k_L+ k_R-u}, ~ t_L{=}\frac{A_R}{A_L}{=}\frac{2k_L}{k_L+ k_R-u}
\end{equation}
and 
\begin{equation}
r_R{=}\frac{A_R}{B_R}{=}\frac{k_R-k_L+u}{k_R+k_L-u}, ~ t_R{=}\frac{B_L}{B_R}{=}\frac{2 k_R}{k_R+k_L-u}, 
\end{equation}
where $u=2\mu V_2/\hbar^2$.
Reflection probabilities    are obtained as $R_L=|r_L|^2$ and $R_R=|r_R|^2$ but  transmission probabilities  for semi-infinite potentials when $E>V_1$ are found as
\begin{equation}
T_L=\frac{k_R}{k_L} |t_L|^2, \quad  T_R=\frac{k_L}{k_R} |t_R|^2 \quad \Rightarrow T_L=T_R.
\end{equation}
One can readily see that $k_L+k_R=u$ is the condition of spectral singularity  at which
 \begin{eqnarray}
\hspace*{-0.2cm}|\det(S(k_L,k_R))|{=}\left|\frac{u{+}(k_L{+}k_R)}{u{-}(k_L{+}k_R)}\right|{=}\infty, \nonumber \\  |\det(S({-}k_L,{-}k_R))| {=} 0
\end{eqnarray}
the modulus of the determinant of time-reversed $S$-matrix vanishes and real energy
turns out to be 
\begin{equation}
E_*= \left(\frac{U^2+V_1}{2 {U}}\right)^2, \quad U=\frac{\sqrt{2\mu}}{\hbar} V_2, \quad  U^2>V_1
\end{equation}
But when $U^2<V_1$ (i.e., $k_L-k_R=u$) 
\begin{equation}
E_i=\left(\frac{U^2+V_1}{2 {U}}\right)^2, \quad r_R(E_i)=0, \quad t_R(E_i)=1.
\end{equation}
Eq. (17) does give a scope for right invisibility of (10) yet it is belied by noting that the transmission probability for right incidence is $T(E_i)=k_L/k_R \ne 1$. So the potential (10) becomes only right-reflectionless at $E=E_i$. Notice that parametric conditions (16,17) of SS and reflectionless are mutually exclusive. CPA can occur even if $V_1=0$, this means that imaginary Dirac delta potential alone (with $V_2>0$) is the simplest model of CPA. Interestingly, here it turns out that  the presence of semi-infinite step potential does not hamper CPA. We claim that the model (10) is the second simplest model of CPA so far. It also has pedagogic advantage. However, the semi-infiniteness of the real part adds novelty in the phenomenon of CPA. 
The analytical demonstration of SS and CPA in (10) has been carried out in Eqs. (12-17) above. For pictorial demonstration,
taking $2\mu=1=\hbar^2$, and $V_1=3, V_2=3$, we present  $T(E), R_L(E)$, and $R_R(E)$  to show spectral singularity at $E=E_*=4$  and the determinant  of the two port time-reversed S-matrix vanishing  at $E=E_*=4$: $|\det(S(-k_L,-k_R ))|=0$ (see Fig. 2).
     
                                                                                                    Next, we consider the potential profile (11) in (1), using the standard transformation [22,23]
\begin{equation}
y{=}\frac{1}{2}[1-\tanh (\frac{x}{2a})],\mbox{and} ~\psi(x){=}y^{-i\beta} (1-y)^{i\alpha} G(y), \end{equation}
we can reduce (1) for (11)
in terms of Gauss Hyper geometric form 
\begin{equation}
y(1-y)\frac{d^2G}{dx^2}+[c-(a+b+1)y]\frac{dG}{dx}-ab G=0,
\end{equation}
where $a=1/2+i(\alpha-\beta+\gamma), b=1/2+i(\alpha-\beta-\gamma), c=1-2i\beta$ and $\alpha=\sqrt{\frac{E}{\Delta}}=k_L a, \beta=\sqrt{\frac{E-V_1}{\Delta}}=k_R a, \gamma=\sqrt{\frac{4iV_2}{\Delta}-\frac{1}{4}}, \Delta=\frac{\hbar^2}{2\mu a^2}.$
This equation has two linearly independent solutions $G_1=~_2F_1(a,b,c,y)$ and $G_2=y^{1-c} ~_2F_1(1+a-c,1+b-c,2-c,y)$. When $x \rightarrow \infty$, $y \rightarrow e^{-x/a}, 1-y \rightarrow 1.$ Also when $x\rightarrow \infty$,$ ~_2F_1(a,b,c,0)=1$ and $~_2F_1(1+a-c,1+b-c,2-c,0)=1.$.
The solutions $G_1$ and $G_2$ give $\psi \sim e^{ik_Rx}$ and $\psi \sim e^{-ik_Rx}$. We choose the first one as it represents a transmitted wave traveling from left to right, finally we seek the solution of (1) for (11) as
\begin{equation}
\psi(x)=N y^{-i\beta} (1-y)^{i\alpha} ~_2F_1(a,b,c,y) \sim N e^{ik_R x}
\end{equation}
Using an identity of Hypergeometric functions as 
\begin{multline}
\hspace*{-.5cm}_2F_1(a,b,c,y){=}\frac{\Gamma(c)\Gamma(c{-}a{-}b)}{\Gamma(c{-}a)\Gamma(c{-}b)}~_2F_1(a,b,a{+}b{-}c{+}1,1{-}y){+}\\
\hspace*{-.56cm}\frac{\Gamma(c)\Gamma(a{+}b{-}c)}{\Gamma(a)\Gamma(b)}(1{-}y)^{c{-}a{-}b}~_2F_1(c{-}a,c{-}b,c{-}a{-}b{+}1,1{-}y) \hspace*{-.5cm}
\end{multline}
When $x\rightarrow -\infty, y \rightarrow 1$, the solution (20) is capable of representing
a linear combination of incident (traveling from left to right)  and reflected waves (traveling right to left) as $x \rightarrow -\infty$
\begin{multline}
\psi(x){\sim}\frac{N'~\Gamma(1{-}2i\beta)\Gamma({-}2i\alpha)e^{ik_Lx}}{\Gamma(1/2{-}i(\alpha+\beta+\gamma))\Gamma (1/2{-}i(\alpha+\beta-\gamma))}\\+\frac{N'~\Gamma(1{-}2i\beta)\Gamma(2i\alpha)e^{{-}ik_Lx}} {\Gamma(1/2{+}i(\alpha{-}\beta{-}\gamma)) \Gamma(1/2{+}i(\alpha{-}\beta{+}\gamma))}
\end{multline}
Equations (20) and (22) help us to get the reflection and transmission  amplitudes for incidence
from left as
\begin{multline}
r(\alpha,\beta){=}\frac{\Gamma(2i\alpha)\Gamma(1/2{-}i(\alpha{+}\beta{+}\gamma))\Gamma (1/2{-}i(\alpha{+}\beta{-}\gamma))}{\Gamma({-}2i\alpha)\Gamma(1/2{+}i(\alpha{-}\beta{-}\gamma)) \Gamma(1/2{+}i(\alpha{-}\beta{+}\gamma))}\\  t(\alpha,\beta){=}\frac{\Gamma(1/2{-}i(\alpha{+}\beta{+}\gamma))\Gamma (1/2{-}i(\alpha{+}\beta{-}\gamma))} { \Gamma(1{-}2i\beta)\Gamma({-}2i\alpha)}.
\end{multline}
Poles or zeros at $\alpha=\pm i n$ or $\beta=-i (n+1)/2$ in (23), where $n=0,1,2,3,...$ are un-physical.
For the  incidence from the right, similarly  we obtain
\begin{eqnarray}
r_L(k_L,k_R)&=&r(\alpha, \beta)~  \mbox{and} ~ t_L(k_L,k_R)=t(\alpha, \beta); \nonumber \\
r_R(k_L,k_R)&=&r(\beta, \alpha)~  \mbox{and} ~ t_R(k_L,k_R)=t(\beta, \alpha)
\end{eqnarray}
leading to non-reciprocity of transmission amplitudes (8)
\begin{equation}
t_L k_R= t_R k_L
\end{equation}
Let $\gamma=q+is, q,s \in {\cal R}$, the second term in numerator of $r_L$ and $t_L$ becomes $\Gamma[1/2-s-i(\alpha+\beta-q)]$, by demanding 
\begin{equation}
1/2-s=-n \quad \alpha+\beta=q,\quad n=0,1,2,3,...,
\end{equation}
we get the real 
\begin{multline}
E_*{=}\left (\frac{W^2 +V_1}{2W}\right)^2,~ V_2{=}\frac{\Delta}{4}[(2n+1)q+i(n(n+1)-q^2)],\\ n\in I^+,~  W=q\sqrt{\Delta}, ~ W^2>V_1,
\end{multline}
the energy of spectral singularity.
Using (23,24), we can write $r_L(-k_L,-k_R)r_R(-k_L,-k_R)=X(\alpha,\beta)/ Y(\alpha, \beta)$ and $t_L(-k_L,-k_R)t_R(-k_L,-k_R)=X(\alpha,\beta)/ Z(\alpha, \beta)$ 
where 
	\begin{equation}
	X(\alpha,\beta)=\frac{(\Gamma(1/2{+}i(\alpha{+}\beta{+}\gamma))\Gamma(1/2{+}i(\alpha{+}\beta{-}\gamma)))^2}{\Gamma(2i\alpha)\Gamma(2i\beta)}.
	\end{equation}
	\begin{multline}
	Y(\alpha,\beta)=\frac{\Gamma(1/2{-}i(\alpha{-}\beta{+}\gamma))\Gamma(1/2{-}i(\alpha{-}\beta{-}\gamma))}{\Gamma(-2i\alpha)} \\ \frac{
		\Gamma(1/2{+}i(\alpha{-}\beta{+}\gamma)) \Gamma(1/2{+}i(\alpha{-}\beta{-}\gamma))}{\Gamma(-2i\beta)},
	\end{multline}
	\begin{equation}
\hspace*{-3.cm}	Z(\alpha, \beta)= \Gamma(1+2i\alpha) \Gamma(1+2i\beta).
	\end{equation}
Under the condition of spectral singularity (26), $X(\alpha,\beta)$ remains a non-zero and finite multiplicative factor in $\det(S(-k_L,-k_R))=X(1/Y-1/Z)$, where 
\begin{multline}
\hspace*{-.3cm}Y^{-1}{=}\frac{\Gamma({-}2i\alpha) \Gamma({-}2i\beta)}{\Gamma(1{+}n{-}2i\alpha) \Gamma({-}n{+}2i\alpha) \Gamma(1{+}n{-}2i\beta) \Gamma({-}n{+}2i\beta)} 
\end{multline}
\vspace{-.9cm}
\begin{multline}
\hspace*{-.3cm}Y^{-1}{=}-\frac{1}{\pi^2} \sinh 2\pi\alpha \sinh 2\pi\beta ~\Gamma(-2i\alpha) \Gamma(-2i\beta)=Z^{-1}
\end{multline}
Leading to $\det(S(-k_L,-k_R))=0$ at $E=E_*$ (26) which is the analytic demonstration of the phenomenon of CPA in (11). For  simplifications in  (31) to the form (32), we have made multiple use of a property of Gamma functions expressed as $\Gamma(1-z)\Gamma(z)=\frac{\pi}{\sin \pi z}$ [19].

Next, when $W^2<V_1$ (i.e., $\alpha- \beta= q, s=n+1/2, V_2$ (as in Eq. (27))
\begin{multline}
E_z{=}\left (\frac{W^2 +V_1}{2W}\right)^2, \quad r_R(E_z){=}0, \quad t_{L,R}(E_z) {\ne} 1, 
\end{multline}
Further, when $W^2<V_1$ and $n=0$ $(V_2=\frac{\Delta}{4}(q-iq^2))$
\begin{equation}
E_i=\left (\frac{W^2 +V_1}{2W}\right)^2, \quad r_R(E_i){=}0, \quad t_R(E_i)= 1,
\end{equation} 
This can be readily checked  when $s=n+1/2$ and $\alpha-\beta=q$ the argument of Gamma function in the denominator of $r_R=r(\beta, \alpha)$ in (23,24) becomes a negative integer, as $\Gamma[-n]=\infty, n=0,1,2,3,...$, we get  
$r_R(E_i)=0$ but $t_R(E_i)=\frac{\Gamma(n+1-2i\alpha) \Gamma(n-2i\beta)}{\Gamma(1-2i\alpha) \Gamma(-2i\beta)}$, which is 1 only when $n=0.$ However,  the result that $T_R(E_i)=k_L/k_R$ (probability of transmission) for incidence from right  would prevent even the right invisibility [14]. So the potential (11) is eventually right-reflectionless where conditions (33,34) are met.
\begin{figure}[h]
	\centering
	\includegraphics[width=7 cm,height=5 cm]{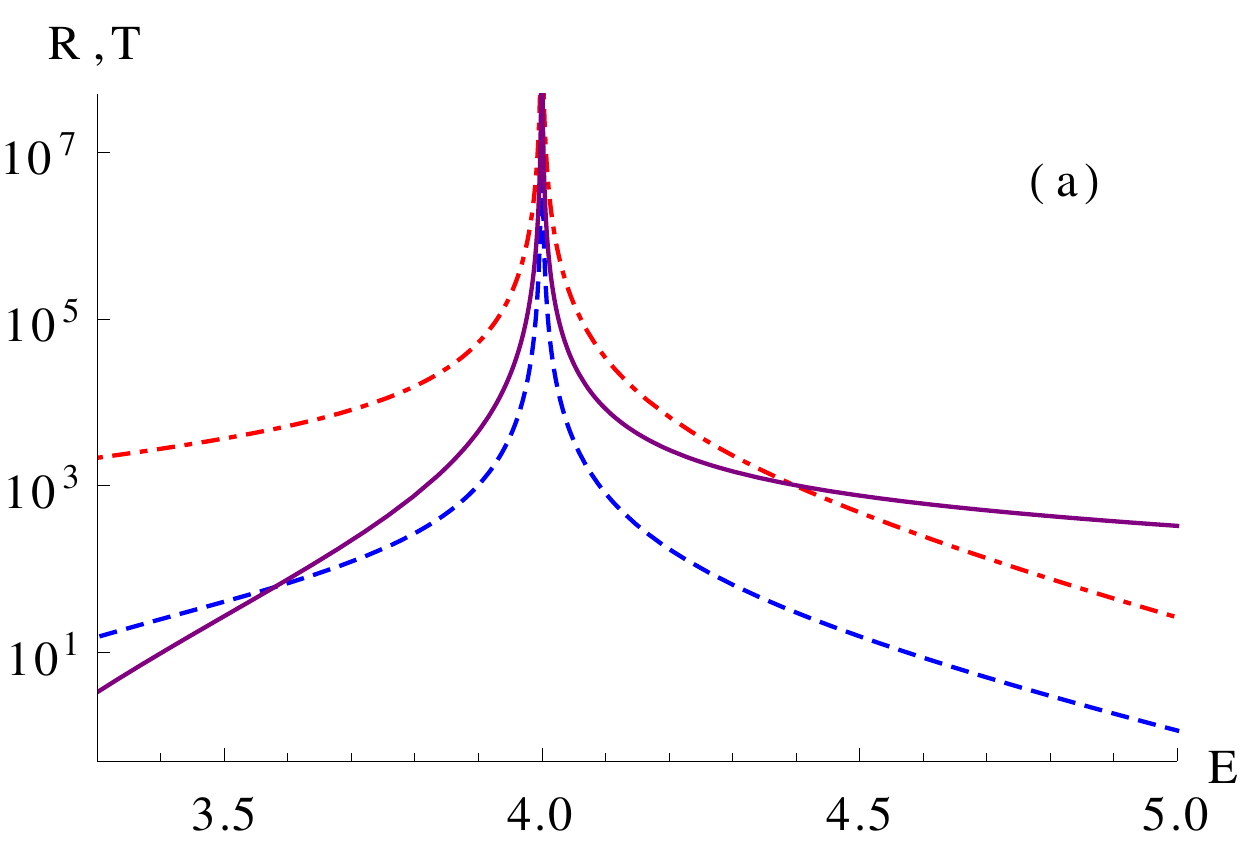}
	\hspace{.5cm}
	\includegraphics[width=7 cm,height=5 cm]{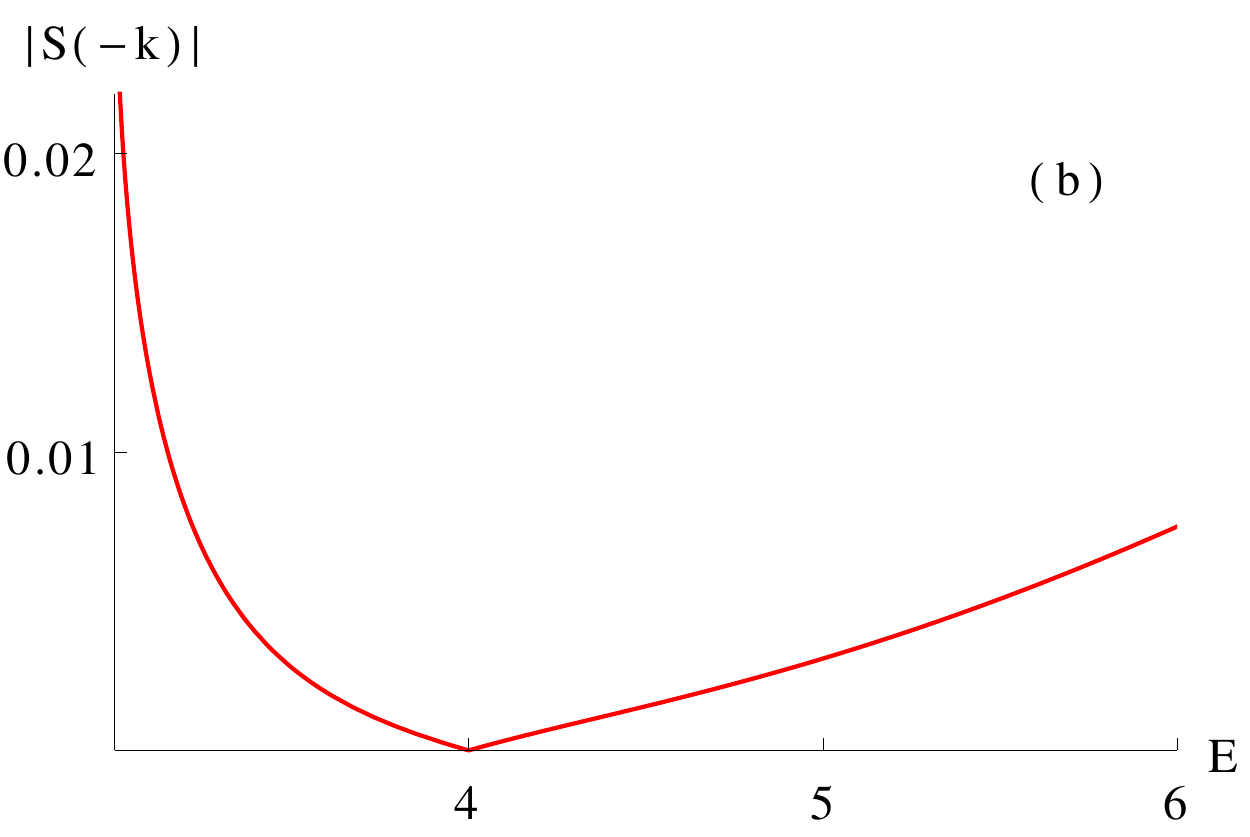}
	\caption{The same as in Fig. 2 for the non-Hermitian version of Eckart potential profile (11). Here, $2\mu=1=\hbar^2$, $V_1=3; q=3, n=1, E_*=4$ (see Eq. (27))}
\end{figure}
Apart from analytic demonstration of SS, CPA and reflectionlessness in the potential (11) through
Eqs. (23-34), in Fig. 3, we present a pictorial demonstration when $V_1=3, q=3, n=1$ yielding $E_*=4$ (26).

The two-port S matrix for coherent scattering derived here (7)  is more general than the one discussed previously [8,11,16,18]. The Eqs. (3-7) are useful for analytically intractable profiles and our conclusions are also based on other profiles such as $V(x)=V_1[1+\mbox{erf}(x/a)]+iV_2 e^{-x^2/a^2}$. We would like to remark that for energies $E<V_1$, $R_L(E)$ or $R_R(E)$ may alone (not $T(E)$) have singularities which can not be admitted as spectral singularity. Some of these unphysical poles have been mentioned below Eq. (23).  

We conclude that the semi-infiniteness of the real part and the non-reciprocity of transmission amplitudes do not hamper the interesting critical phenomena of spectral singularity, coherent perfect absorption and one-sided reflectionlessness, they occur yet again. All of these occur at energies $E>V_1$ and when ${\cal R} (V_2)>0.$ The first two phenomena require the strength of the imaginary part of the potential to be larger. We find that one sided reflectionlessness can occur for lesser values of  ${\cal R} (V_2)>0.$ Very interestingly, the invisibility
gets ruled out despite $r(E_i)=0$ and $t(E_i) =1$ (but $T(E_i)\ne 1$) on one side of these semi-infinite models (10,11).
It may be remarked that the (non)occurrence of invisibility is cumbersome and difficult to detect  as  experienced in Ref.[15]. However, here this could be done easily. Let us call the situation of $r(E_z)=0$ and $T(E_z) \ne 1$ on one side as one-sided reflectionlessness. We find that  existence of one sided reflectionlessness and spectral singularity are mutually exclusive for a fixed semi-infinite  non-Hermitian potential. This is an other distinctive feature of semi-infinite medium. The similarity of the results in Eqs. (16,27) for two potentials (10,11) is tantalizing.

Investigations of coherent injection at non-Hermitian mediums have been throwing interesting 
surprises and revealing novel phenomena in the recent past. In this scenario, our proposal of semi-infinite optical potentials provides a new avenue. We hope that the two proposed exactly solvable models and their explicit results which are surprisingly simple  will be found useful in both theory and experiments.

\end{document}